\documentclass{ws-procs11x85}
\usepackage{ws-procs-thm}

\begin{document}
\title{ZetaDial: dialing net charge of protein binders at inference time
for therapeutic developability}

\author{Mohammed Sameer Syed$^{1}$ and Tamara Dinneen$^{2}$}

\address{$^{1}$E-mail: mohammedsameersyed1@gmail.com\\
$^{2}$E-mail: tamaradinneen@gmail.com}

\begin{abstract}
Net charge is a developability-relevant property of therapeutic binders, associated
with viscosity, clearance, nonspecific interaction and aggregation, and antibody
screens already use charge-related criteria\cite{sharma2014insilico,jain2017biophysical,raybould2019tap,igawa2010reduced}. Yet inverse-folding pipelines expose no
way to set it to a target value.
ProteinMPNN exposes global and per-residue amino-acid biases, and BindCraft offers
weight choices and custom losses, but neither supplies a built-in per-protein
feedback loop that measures realised charge after sampling and corrects it to a
requested setpoint.\cite{dauparas2022proteinmpnn,pacesa2025bindcraft} Other
inference-time methods now target charge explicitly, including Feynman--Kac steering;\cite{fk2025steering} ZetaDial contributes a post-sampling, per-protein
secant controller around fixed-backbone ProteinMPNN. On matched stochastic
benchmarks, the secant loop reduced mean absolute error relative to a fixed-slope
per-protein loop on RCSB complexes ($5.17$ versus $6.46$ charge units) and Cas13
monomers ($5.57$ versus $8.23$). Relative to the optimised matched global $\beta$, it
reduced RCSB error from $11.71$ to $5.17$ (cluster bootstrap $p<0.001$) and was
statistically indistinguishable on Cas13 ($5.47$ versus $5.57$). In $800$ overlapping eight-protein subsets pooled
from those two datasets, sensitivity heterogeneity was associated with calibration
gain (Pearson $r=0.79$); this is descriptive resampling, not an independent
prospective decision rule. Foldability deteriorated as $|\beta|$ increased, most
strongly at $\pm3$ and substantially for Cas13 at $\pm1.5$. In the full
$52$-complex seed-$0$ analysis, reference-based DockQ declined clearly at $\pm3$
but not at $\pm1.5$. Five new seeds on eight complexes selected for acceptable
seed-$0$, $\beta=0$ DockQ showed paired declines at every nonzero setting; this
selected replication does not estimate the effect size for all $52$ complexes. In
exploratory BindCraft sweeps, PD-L1 designs moved towards near-neutral charge while
retaining similar maximum i\_pTM values, but success-rate confidence intervals
overlapped; IL-7R$\alpha$ responses were non-monotonic, and RBD produced no strong
designs. A fixed-backbone C$\alpha$-neighbour analysis reported smaller same-sign
charge-patch proxies near neutral charge, but this proxy is not a measured
electrostatic surface or experimental developability endpoint. 

\end{abstract}


\section{Introduction}\label{sms:intro}

Inverse folding, predicting a sequence that folds to a given backbone, has been
transformed by deep generative models led by
ProteinMPNN,\cite{dauparas2022proteinmpnn} and de-novo binder pipelines such as
BindCraft,\cite{pacesa2025bindcraft} built on RFdiffusion\cite{watson2023rfdiffusion}
and AlphaFold2,\cite{jumper2021alphafold} now compose these foundation models into
functional binders in one shot, an instance of generative therapeutic design. Yet
they are \emph{unconditioned} on the biophysical properties that decide a molecule's
fate downstream, the gap between raw generative output and a clinically viable
therapeutic.

Net charge is one developability-related descriptor. In therapeutic
antibodies, variable-domain charge and charge distribution have been associated
with clearance, self-association and high-concentration viscosity, while extreme
values can accompany polyspecificity or aggregation
risk.\cite{sharma2014insilico,jain2017biophysical,raybould2019tap,igawa2010reduced,schoch2015charge,dattamannan2021charge,kelly2015cross,deepsp2024}
Published studies have proposed numerical screening windows for particular
antibody datasets, but those ranges are empirical heuristics rather than universal
go/no-go criteria for all protein binders. Existing inverse-folding and
binder-design tools can influence composition through biases, model weights or
custom objectives, but they do not provide the post-sampling per-protein secant
controller evaluated here.

We treat charge as a continuously tunable design variable without retraining and
carry that control into a binder-design workflow. ProteinMPNN already provides
global and per-residue amino-acid biases, BindCraft exposes MPNN weight choices and
permits custom losses, and recent methods steer charge or pI using analytic bias,
position-specific bias or target-based inference-time
rewards.\cite{raghavan2026protnhf,deng2025epitope,fk2025steering} ZetaDial differs
in a testable way: after each fixed-backbone ProteinMPNN sample, it
measures the realised scalar charge and uses a guarded secant update to select the
next bias. To our knowledge, prior work has not evaluated this particular
post-sampling controller against matched global-bias and fixed-slope comparators
while carrying the resulting sequences into a binder pipeline with predicted
interface metrics tracked.

\begin{arabiclist}
\item \textbf{A post-sampling, per-protein charge controller with a guarded
adaptive-slope update.} Among the two per-protein feedback controllers tested, the
secant loop had lower mean MAE on both benchmark datasets: $5.17$ versus $6.46$ on
RCSB complexes and $5.57$ versus $8.23$ on Cas13 monomers. Relative to the matched
global bias, it was lower on the heterogeneous RCSB set and indistinguishable on
Cas13.
\item \textbf{A descriptive analysis of when per-protein calibration may help,
plus a composition-based sensitivity model.} Across $800$ overlapping
eight-protein subsets resampled from the two benchmark datasets, sensitivity
heterogeneity was associated with observed calibration gain ($r=0.79$). Grouped
cross-validated $R^2$ values of $0.77$ and $0.53$ show that composition predicts
part of the sensitivity variation, suggesting a possible warm-start strategy that
remains to be tested prospectively.
\item \textbf{A characterisation of the charge - foldability trade-off using
isolated-chain self-consistency and complex-aware AlphaFold2-Multimer predictions.}
Foldability declined with stronger bias. In the full $52$-complex seed-$0$ cohort,
ipSAE declined at every nonzero setting, whereas DockQ declined clearly at
$\pm3$. In an independent-seed replication on eight baseline-quality-selected
complexes, overall DockQ declined at all four nonzero settings. Fold-tolerant
conditional subsets were too small and metric-dependent to support an interface
preservation claim.
\item \textbf{Exploratory transfer into BindCraft, with predicted binding metrics
and computed developability proxies across three targets.} The sweeps show the charge bias changes realised binder charge, but binding-score patterns
differ by target and sample sizes do not establish universal or
target-specific tolerance windows. A fixed-backbone C$\alpha$-neighbour proxy is
reported separately from experimental developability endpoints.
\end{arabiclist}
On the held-out data, a geometric proxy counted more putative salt bridges in
ProteinMPNN designs than in native structures ($11.3$ versus $9.1$,
$n\approx170$, $p=7\times10^{-5}$). This single proxy supports neither a global
claim that charge placement is better than native nor a developability conclusion;
it is reported only to distinguish charge controllability from an assertion that
ProteinMPNN is generally deficient.

\section{Related Work}\label{sms:related}

\paragraph{Controllable and guided generation.}
Classifier guidance, biased sampling and inference-time steering can target protein
properties without retraining. ProtNHF\cite{raghavan2026protnhf} provides smooth
sampling-time control over net charge; position-specific bias has been used to
influence pI;\cite{deng2025epitope} and Feynman--Kac
steering\cite{fk2025steering} explicitly evaluates a reward that penalises
deviation from a target charge $Q^\star$ while resampling particles.
Predictor-guided methods such as ProteinGuide\cite{proteinguide} likewise condition
generation on requested property values. These methods implement different forms
of target-based steering. ZetaDial's distinct scope is a post-sampling,
fixed-backbone ProteinMPNN loop that measures realised charge after each design and
updates a scalar bias by a guarded secant rule. Its matched global-$\beta$
comparator evaluates the residual left by one open-loop bias choice, not the full
class of prior guidance methods.

\paragraph{Biased and retrained ProteinMPNN.}
ProteinMPNN provides global and per-residue bias inputs. BindCraft can switch
between soluble and original MPNN weights and allows custom losses; recent
pipelines have also used surface substitutions, charge-related losses, retraining
or preference alignment to influence composition and
developability.\cite{deng2025epitope,goverde2024soluble,hou2026protalign} We did
not identify a prior matched evaluation of the exact guarded secant controller used
here around fixed-backbone ProteinMPNN. That algorithmic and empirical distinction
is narrower than claiming that no prior method offers a signed charge target.

\paragraph{Charge, foldability, and interface electrostatics.}
Protein supercharging established that surface net charge can be pushed tens of
units without abolishing the fold, with a soft ceiling where solubility
degrades;\cite{liu2007supercharge,der2013supercharge} increased negative surface
charge correlates with solubility while positive patches predict
insolubility,\cite{chan2013soluble,kramer2012solubility} explaining the asymmetric
tolerance we observe. In binder design, recent complementarity-based pipelines
optimize \emph{shape} complementarity (e.g.\ HECTOR,\cite{maksymenko2025hector}
which reaches nanomolar binders, including against IL-7R$\alpha$), and a large
meta-analysis\cite{overath2025binders} finds interface \emph{electrostatic}
complementarity only a weak predictor of success, while the nearest charge-control
work imposes interface electrostatics for pH-switching\cite{baker2025phbinders}
rather than a net-charge setpoint. Net charge as a calibrated, tunable handle on
the designed binder is thus largely unexplored; it is what we add to this pipeline.

\section{Methods}\label{sms:methods}

\paragraph{Steering primitive (a native feature).}
ProteinMPNN samples autoregressively from per-position logits and exposes a
per-amino-acid additive bias; the equivalent input exists in ColabDesign's
MPNN.\cite{colabdesign} Our primitive sets that bias with a single signed scalar
$\beta$, adding $\beta$ to the lysine and arginine (K, R) logits and subtracting
$\beta$ from the aspartate and glutamate (D, E) logits at every position. Here
$\beta>0$ pushes designs more positive and $\beta=0$ recovers vanilla ProteinMPNN. We sample at temperature $0.1$; because softmax divides logits by the
temperature, the effective strength of a given $\beta$ scales inversely with it,
which is why the usable $\beta$ range differs between standalone runs and BindCraft.
Net charge is scored as $(\#\mathrm{K}+\#\mathrm{R})-(\#\mathrm{D}+\#\mathrm{E})$;
developability descriptors (net charge at pH~$7.4$, isoelectric point) additionally
account for histidine and chain termini via a Henderson--Hasselbalch calculation.
The two agree closely: across our designs the count and the pH-$7.4$ net charge
differ by $0.21$ charge units on average (median $0.17$, maximum $0.50$), well below
the control resolution, so calibrating on the count effectively calibrates the
developability-relevant pH-$7.4$ charge.

\paragraph{Closed-loop calibration (ZetaDial).}
Because a fixed $\beta$ produces a protein-dependent charge shift, hitting a
\emph{specified} target requires per-protein calibration. For each protein we probe
with single samples across a small $\beta$ grid, estimate the local slope
$\mathrm{d}q/\mathrm{d}\beta$ by central difference (floored at a small positive
value to avoid overshoot on charge-insensitive proteins), initialise
$\beta=(q^\star-q_0)/\mathrm{slope}$, sample, measure the miss, and correct, for a
fixed budget of three iterations, keeping the closest sample. The charge response is
\emph{saturating}: a fixed slope estimated near $\beta=0$ over-predicts the effect of
large $\beta$, so a naive proportional loop overshoots, most damagingly on
charge-insensitive monomers. ZetaDial therefore uses an adaptive-slope (secant)
update: after each new sample it re-estimates the local slope from the two most
recent $(\beta,q)$ evaluations, guarded to update only when the response is locally
monotone and otherwise falling back to the probe slope. This tracks the curvature of
the response and converges in the same three-sample budget. All sampling is at
temperature $0.1$ with recorded seeds, and a single design is drawn per probe and per
iteration, so the reported errors already fold in sampling noise rather than
averaging it away; averaging two samples per evaluation further widens the secant
advantage. As a matched baseline we select the lowest-error single global $\beta$
across all proteins for each target and report the residual it leaves; this is an
open-loop global-bias comparator of the same class as analytic monotonic-knob
steering.\cite{raghavan2026protnhf} For an internally consistent comparison we run
all three controllers, global-$\beta$, fixed-slope loop, and ZetaDial's secant loop,
on a \emph{single} set of absolute charge-shift targets, scoring each protein only
where the target is reachable, so Table~\ref{sms:tbl-control} is a like-for-like
head-to-head under real sampling.

\paragraph{Predicting charge sensitivity.}
A protein's sensitivity $\mathrm{d}q/\mathrm{d}\beta$ governs how strong a bias a
target demands. We fit a ridge regressor using length and titratable-residue counts
and fractions, evaluated by grouped cross-validation within each dataset. The
resulting prediction is evaluated here as a sensitivity estimate; using it as a
probe-free initialisation for the feedback loop is a proposed application rather
than a separately benchmarked controller in this study.

\paragraph{Datasets and metrics.}
We evaluate on (1) charged protein--protein complexes from the RCSB, split by $30\%$
sequence-identity clustering with MMseqs2\cite{steinegger2017mmseqs2} so that no
test protein is a homolog of any training protein, and (2) $96$ de-novo Cas13
binder monomers. For the RCSB complexes the controller targets the complex-level net
charge summed over all chains, which is why the reachable span reaches $\sim$915
units on the largest multi-chain complexes; the Cas13 monomers and the BindCraft
binders are single chains, matching the therapeutic scenario in which only the binder
is designed. Foldability is measured by ESMFold\cite{lin2023esmfold}
self-consistency RMSD (scRMSD).

\paragraph{Complex-aware refolding and interface scoring.}
We refolded $52$ RCSB complexes at $\beta\in\{-3,-1.5,0,1.5,3\}$ with
AlphaFold2-Multimer, using one recorded seed per complex and setting. The
pre-specified prediction-only interface endpoint was ipSAE. We additionally scored
each prediction against its native RCSB complex with DockQ v2.1.3,\cite{mirabello2024dockqv2}
using the recorded native-to-model chain map. Because the designed chain contains
intentional substitutions, identity-only residue matching would omit redesigned
positions. For the primary DockQ score, residue labels were therefore normalised
at sequence-aligned corresponding positions before scoring; coordinates and atom
records were unchanged, all model residues were mapped, and the unmodified
identity-only score was retained as a sensitivity analysis. We then selected eight
complexes whose seed-$0$, $\beta=0$ DockQ was at least $0.23$ and repeated all five
settings with seeds $1$--$5$ ($200$ additional predictions). This is a
baseline-quality-selected stochastic replication, not a representative subsample
of the $52$ complexes. Changes are paired to $\beta=0$ within complex and seed.
Confidence intervals use $10{,}000$ bootstrap resamples clustered by complex.
Fold-tolerant subsets retain predictions whose designed-chain pLDDT is within $10$
points of the paired baseline.

BindCraft\cite{pacesa2025bindcraft} designs binders
by AlphaFold2 back-propagation, ProteinMPNN redesign, and
AlphaFold2-multimer\cite{evans2021multimer} re-prediction with Rosetta scoring; we
inject the identical charge bias into its MPNN step, attending to that
implementation's amino-acid ordering (AlphaFold's restype alphabet, which differs
from raw ProteinMPNN's; indexing with the wrong order silently biases the wrong
residues and moves no charge). We design against PD-L1 (BindCraft example),
IL-7R$\alpha$ (PDB 3DI2 chain B), and SARS-CoV-2 RBD (PDB 6M0J chain E), each across
a per-target $\beta$ grid. The bias is applied uniformly to all designed positions;
because BindCraft holds the interface residues fixed during redesign, it acts on
non-interface (largely solvent-exposed) positions by construction, and restricting it
to explicitly surface-exposed residues is a straightforward extension. Filters are
disabled so every design is recorded; per
$(\mathrm{target},\beta)$ we report achieved net charge, maximum interface pTM
(i\_pTM), and a success rate (fraction of designs with i\_pTM $\ge0.5$), which are
robust to weak designs that mean i\_pTM conflates with target difficulty.

\section{Results}\label{sms:results}

\subsection{Secant feedback reduces mean charge-hit error under heterogeneity}
The raw bias moves net charge smoothly and monotonically; closing the loop turns it
into a controller that hits specified targets. On one matched set of absolute
charge-shift targets, scored per protein wherever reachable and under real
stochastic sampling, we compare three controllers head to head
(Table~\ref{sms:tbl-control}). Two findings stand out.

First, among the two per-protein feedback controllers tested, the secant loop had
lower mean MAE on both datasets: $5.17$ versus $6.46$ on RCSB complexes and $5.57$
versus $8.23$ on Cas13 monomers. Cluster-bootstrap differences for secant minus
fixed were $-1.28$ [$-2.38$, $-0.40$] and $-2.67$ [$-3.17$, $-2.16$],
respectively. The result concerns mean error: on RCSB the secant loop was closer in
$42.9\%$ of individual cells, indicating that its mean advantage largely reflects
reduction of the fixed loop's overshoot tail rather than uniform per-target
dominance.

Second, per-protein calibration improved on a single matched global bias in the
heterogeneous RCSB set but not in the Cas13 set. On RCSB, the secant loop reduced
MAE from $11.71$ to $5.17$, with a clustered secant-minus-global difference of
$-6.53$ [$-8.03$, $-5.04$]. On Cas13, global $\beta$ and the secant loop had MAE
of $5.47$ and $5.57$, respectively, with a difference of $+0.10$ [$-0.53$,
$+0.73$]. Unbiased filtering using the minimum-error design among $20$ samples
produced substantially larger errors,
$26.7$ on RCSB and $29.9$ on Cas13, in this matched target grid. These results
bound the claim to the tested datasets, targets and sampling budget.

\begin{figure}[h]
\centerline{\includegraphics[width=4.0in]{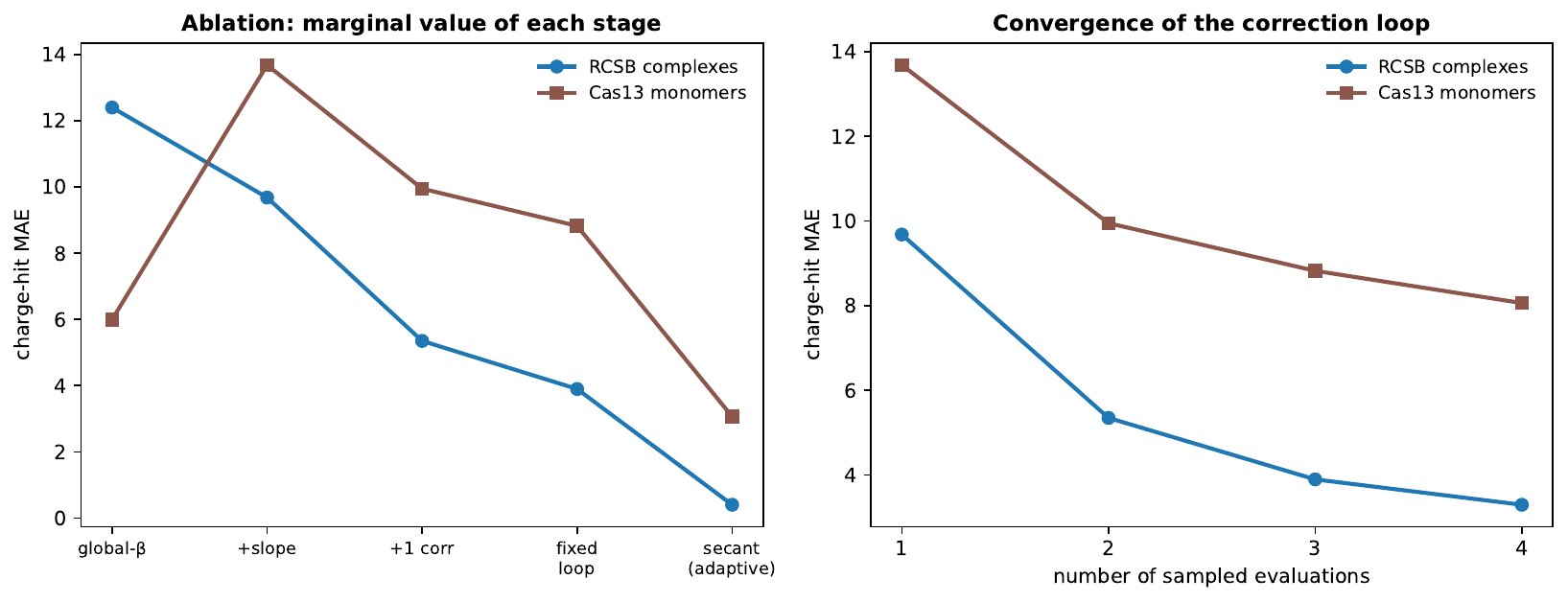}}
\caption{Marginal value of controller stages on each protein's mean
charge-response curve. These deterministic mean-curve errors are optimistic
relative to the single-sample values in Table~\ref{sms:tbl-control}. Left: global
$\beta$, per-protein slope, one correction, the fixed-slope loop and the adaptive
secant loop. The secant stage has the lowest mean error in the displayed ablation,
while matched stochastic cells show heterogeneous per-target outcomes. Right: mean
error versus number of sampled evaluations.}
\label{sms:fig-ablation}
\end{figure}

The stagewise ablation in Fig.~\ref{sms:fig-ablation} uses each protein's mean
charge-response curve, rather than deployable single-sample outcomes. On those mean
curves, adding a per-protein slope, correction steps and the adaptive update reduced
mean error. In the matched stochastic cells, the secant-minus-fixed
cluster-bootstrap differences were $-1.28$ [$-2.38$, $-0.40$] on RCSB and $-2.67$
[$-3.17$, $-2.16$] on Cas13. Because the RCSB fixed loop was closer in a slight
majority of individual cells, the result should be interpreted as lower mean loss
through control of large overshoots, not superiority on every target. Achieved and
target charge were correlated ($r=0.95$), but band-hit performance should be
reported directly if landing inside a developability window is the operative
endpoint.

\begin{table}[h]
\tbl{Charge-hit error on one matched set of absolute charge-shift targets under
ProteinMPNN sampling, scored per protein where reachable (charge units; lower is
better; one sample per evaluation). Columns count $912$ RCSB and $768$ Cas13
(protein, target) cells. The secant loop had lower mean MAE than the fixed-slope
loop in both datasets, lower MAE than global $\beta$ on RCSB, and similar MAE to
global $\beta$ on Cas13. Minimum-error-of-$20$ unbiased filtering had higher MAE in this
target grid.}
{\begin{tabular}{@{}lrr@{}}\toprule
Controller & RCSB complexes ($n=912$) & Cas13 monomers ($n=768$)\\ \colrule
Minimum-error-of-$20$ unbiased filtering & $26.7$ & $29.9$\\
Optimised matched global $\beta$ (open-loop) & $11.7$ & $\mathbf{5.5}$\\
Per-protein fixed-slope loop            & $6.5$  & $8.2$\\
ZetaDial (per-protein secant loop)      & $\mathbf{5.2}$ & $5.6$\\ \botrule
\end{tabular}}
\label{sms:tbl-control}
\end{table}

\subsection{Sensitivity heterogeneity is associated with calibration gain in
resampled subsets}
\begin{figure}[h]
\centerline{\includegraphics[width=2.6in]{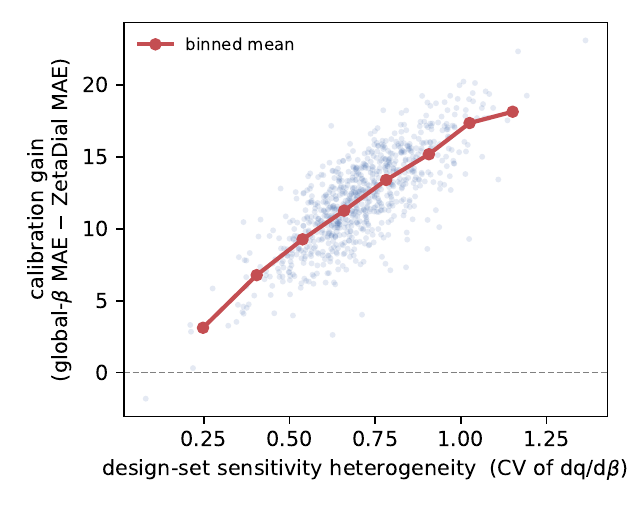}}
\caption{Across $800$ overlapping eight-protein subsets resampled from the two
benchmark datasets, sensitivity heterogeneity was associated with observed
calibration gain (Pearson $r=0.79$). Points are resampled subsets and the line is
the binned mean. Reuse of proteins across subsets means this is a descriptive
mechanism analysis rather than an independent test of predictive performance.}
\label{sms:fig-hetlaw}
\end{figure}

Across $800$ overlapping eight-protein subsets resampled without replacement from
the pooled RCSB and Cas13 benchmark proteins, the coefficient of variation of
charge sensitivity was associated with the observed difference between
global-$\beta$ and secant-loop MAE (Pearson $r=0.79$;
Fig.~\ref{sms:fig-hetlaw}). Because proteins recur across subsets and all subsets
come from only two source datasets, the points are not independent validation
cohorts. The analysis is consistent with the proposed mechanism that a single
global bias leaves larger residuals when responses differ, but it does not by
itself let a practitioner predict prospective gain with calibrated uncertainty.

\subsection{Charge sensitivity is partly predictable from composition and size}
\begin{figure}[h]
\centerline{\includegraphics[width=2.5in]{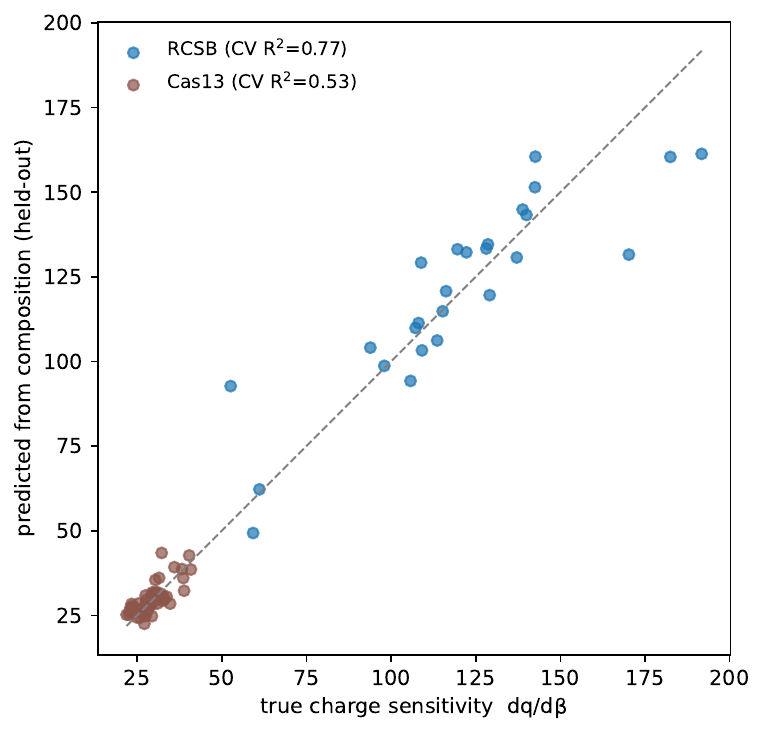}}
\caption{Predicted versus measured charge sensitivity
$\mathrm{d}q/\mathrm{d}\beta$ using composition and size features. Grouped
cross-validated $R^2$ is $0.77$ on RCSB complexes and $0.53$ on Cas13 monomers.
The estimates could be tested as controller initial values; this figure does not
evaluate warm-start performance.}
\label{sms:fig-sensitivity}
\end{figure}

Charge sensitivity $\mathrm{d}q/\mathrm{d}\beta$ was partly predictable from length
and titratable-residue composition, with grouped cross-validated $R^2$ of $0.77$ on
RCSB complexes and $0.53$ on Cas13 monomers
(Fig.~\ref{sms:fig-sensitivity}). Because the response saturates and the Cas13
model leaves substantial unexplained variation, these estimates do not replace
feedback. They could be evaluated as probe-free initial values in a prospective
warm-start comparison; the current results do not measure whether doing so reduces
samples or improves final charge-hit error.

\subsection{Foldability declines as the magnitude of charge bias increases}
\begin{figure}[h]
\centerline{\includegraphics[width=3.8in]{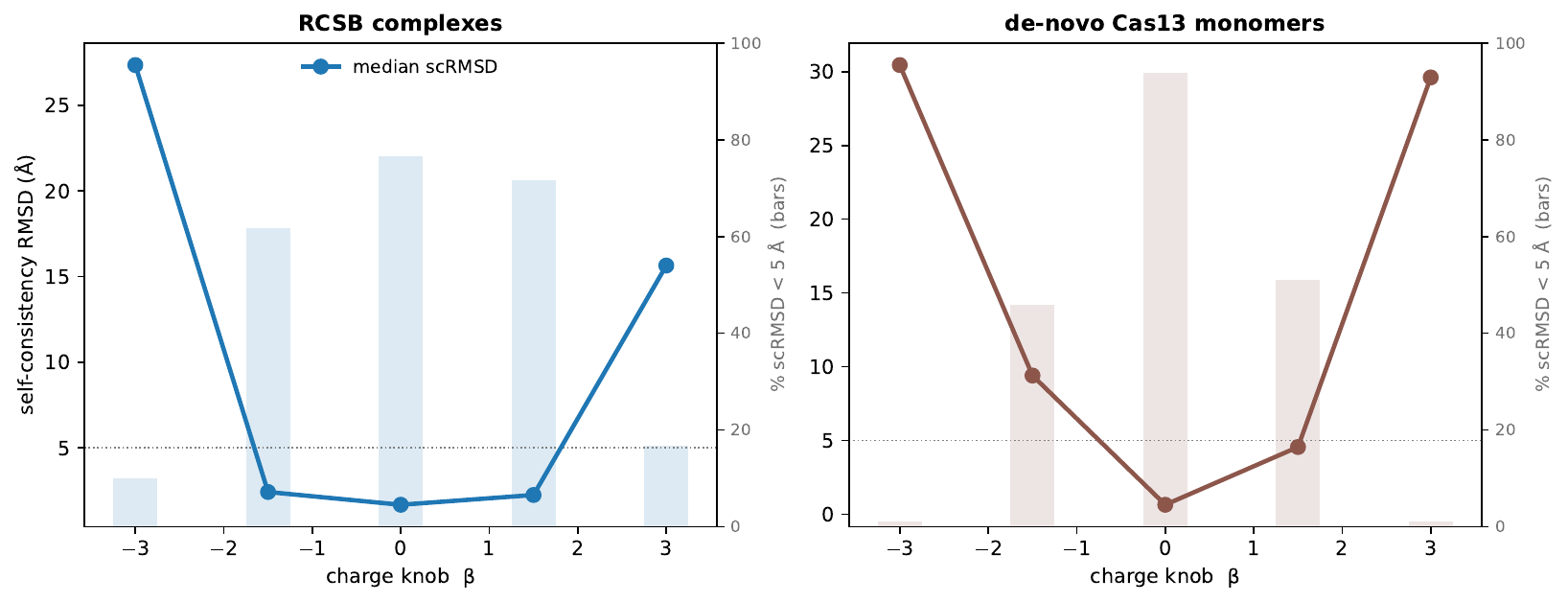}}
\caption{Charge-versus-foldability results for RCSB complexes (left) and de novo
Cas13 monomers (right). Median scRMSD rises and the proportion below $5$\,\AA\
falls as $|\beta|$ increases. The largest losses occur at $\pm3$, while Cas13 also
shows substantial loss at $\pm1.5$.}
\label{sms:fig-foldboth}
\end{figure}

Foldability deteriorated as the charge bias moved away from zero
(Fig.~\ref{sms:fig-foldboth}). For RCSB complexes, $\beta=0$ had median scRMSD
$1.69$\,\AA\ with $76.7\%$ below $5$\,\AA; at $\beta=-1.5$ and $+1.5$ the
medians were $2.43$ and $2.26$\,\AA\ and pass rates were $61.7\%$ and $71.7\%$.
At $\beta=-3$ and $+3$, medians rose to $27.36$ and $15.65$\,\AA\ and pass rates
fell to $10.0\%$ and $16.7\%$. Cas13 monomers were more sensitive: $\beta=0$ had
median $0.64$\,\AA\ and a $93.8\%$ pass rate, while $\beta=-1.5$ and $+1.5$ had
medians of $9.40$ and $4.56$\,\AA\ and pass rates of $45.8\%$ and $51.0\%$; the
extreme settings had median scRMSD near $30$\,\AA\ and about $1\%$ passing.
ESMFold self-consistency folds chains in isolation, so the complex-aware analysis
below addresses a different question and does not remove these designability
losses.

\subsection{Complex-aware refolding shows associated fold and interface
changes}\label{sms:refold}
We refolded $52$ designed complexes across the five charge settings with
AlphaFold2-Multimer; all $52$ returned at every $\beta$. In the full cohort,
designed-chain pLDDT fell at every nonzero setting. The primary interface endpoint,
ipSAE, also declined: mean changes versus $\beta=0$ were $-0.151$ [$-0.238$,
$-0.072$] at $-3$, $-0.056$ [$-0.119$, $-0.001$] at $-1.5$, $-0.093$ [$-0.155$,
$-0.038$] at $+1.5$ and $-0.119$ [$-0.193$, $-0.053$] at $+3$. Reference-based
DockQ showed clear mean losses at the extremes: $-0.116$ [$-0.198$, $-0.034$] at
$-3$ and $-0.115$ [$-0.187$, $-0.045$] at $+3$. At $-1.5$ and $+1.5$, its
intervals included zero: $-0.039$ [$-0.104$, $+0.027$] and $-0.024$ [$-0.103$,
$+0.057$], respectively. Only $11/52$ seed-$0$, $\beta=0$ predictions met the
DockQ acceptable-or-better threshold of $0.23$, so the full-cohort DockQ analysis
is partly floor-limited.

\begin{figure}[h]
\centerline{\includegraphics[width=4.6in]{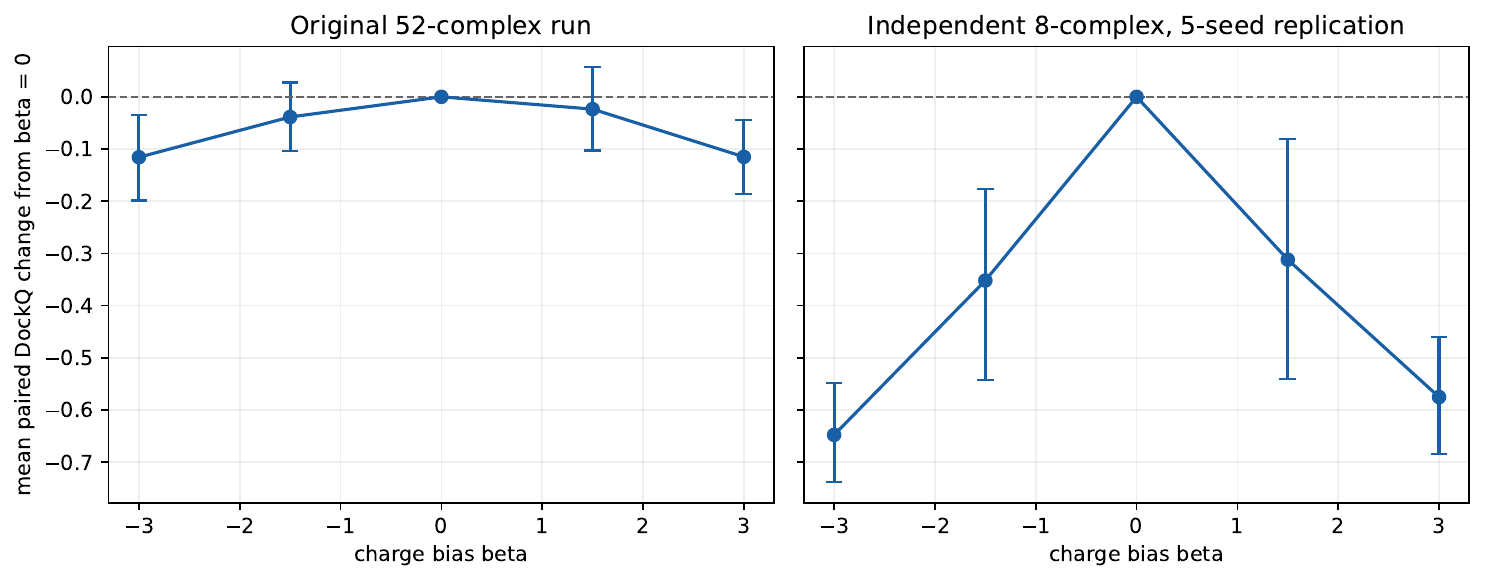}}
\caption{Paired DockQ change from the matched $\beta=0$ prediction. Points are
complex-clustered means and bars are $95\%$ cluster-bootstrap intervals. The
right-hand panel contains eight complexes selected for seed-$0$, $\beta=0$ DockQ
$\geq0.23$ and five new seeds per setting. Its effect sizes should not be
generalised to full $52$-complex cohort or compared between panels as if
cohorts were exchangeable.}
\label{sms:fig-dockq}
\end{figure}

The independent-seed replication used eight complexes selected for acceptable
seed-$0$, $\beta=0$ DockQ. Across five new seeds, mean paired DockQ changes were
$-0.648$ [$-0.738$, $-0.548$] at $-3$, $-0.352$ [$-0.542$, $-0.177$] at $-1.5$,
$-0.312$ [$-0.541$, $-0.080$] at $+1.5$ and $-0.575$ [$-0.684$, $-0.460$] at
$+3$ (Fig.~\ref{sms:fig-dockq}). Every seed had a negative mean change at every
nonzero setting. Across the $40$ complex-by-setting cells, the new-seed mean
profile correlated with seed $0$ at $r=0.988$ and had mean absolute difference
$0.029$; this is descriptive because the selection used seed-$0$ DockQ.
Acceptable-or-better predictions decreased from $39/40$ at $\beta=0$ to $26/40$
at $-1.5$, $19/40$ at $+1.5$, $0/40$ at $-3$ and $5/40$ at $+3$.

Conditioning on designed-chain pLDDT within $10$ points of baseline left only
$3$ complexes ($11$ predictions) at $-1.5$, $4$ ($18$) at $+1.5$, none at $-3$
and one ($5$) at $+3$. At $\pm1.5$, conditional DockQ intervals included zero,
whereas conditional ipSAE intervals did not. The single-complex $+3$ result has no
meaningful between-complex uncertainty. These sparse, metric-dependent conditional
results do not establish that the interface is preserved.

\subsection{Exploratory BindCraft sweeps show different predicted responses by
target}
\begin{figure}[h]
\centerline{\includegraphics[width=4.6in]{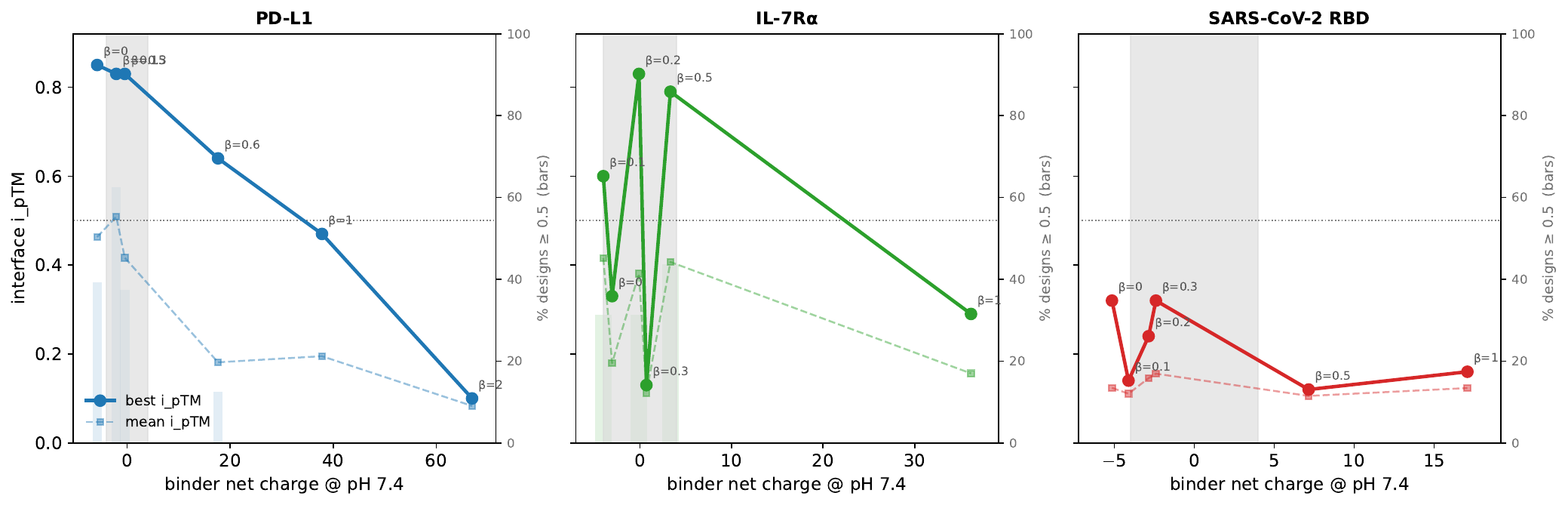}}
\caption{Exploratory BindCraft sweep: maximum and mean interface i\_pTM and the
proportion of designs with i\_pTM $\ge0.5$ versus achieved charge. The shaded
range is an antibody-derived charge heuristic, not a validated cut-off for these
de novo binders. PD-L1 shows similar maximum i\_pTM near neutral charge with
overlapping success-rate uncertainty; IL-7R$\alpha$ responses are non-monotonic;
RBD produces no designs above the chosen i\_pTM threshold.}
\label{sms:fig-tol}
\end{figure}

The charge bias changed realised binder charge in all three BindCraft sweeps, but
the small-step response was not monotonic on every target. PD-L1 and RBD moved
progressively positive across the saved grids. IL-7R$\alpha$ moved from mean charge
$-3.03$ at $\beta=0$ to $-3.99$ at $\beta=0.1$ before becoming positive at larger
settings. Predicted binding metrics also varied by target
(Fig.~\ref{sms:fig-tol}; Table~\ref{sms:tbl-bindcraft}), so these sweeps are
treated as exploratory response profiles rather than calibrated tolerance windows.

For PD-L1, $\beta=0$ produced mean net charge $-5.8$, maximum i\_pTM $0.85$ and a
$39.3\%$ success rate. At $\beta=0.15$, the mean charge was $-2.1$, maximum
i\_pTM was $0.83$ and success was $62.5\%$; at $\beta=0.3$, mean charge was
$-0.5$, maximum i\_pTM was $0.83$ and success was $37.5\%$. The bootstrap success
intervals for baseline [$21.4\%$, $57.1\%$] and $\beta=0.15$ [$41.7\%$, $79.2\%$]
overlap, and no equivalence margin was specified, so these data do not establish
no cost to binding. More positive settings reduced the reported binding metrics,
with $\beta=2.0$ yielding mean charge $+66.9$ and maximum i\_pTM $0.10$. For
IL-7R$\alpha$, baseline produced no designs above i\_pTM $0.5$ and maximum i\_pTM
$0.33$; biased settings produced non-monotonic hit rates, including $31.3\%$ at
$\beta=0.2$ and $43.8\%$ at $\beta=0.5$, followed by $0\%$ at $\beta=1.0$.
These are newly appearing predicted hits rather than preservation of baseline
binders. For RBD, no setting produced a design above i\_pTM $0.5$ and the maximum
was $0.32$, so only charge shifting, not binding preservation, was observed. The
three sweeps suggest different computational response patterns, but their small and
unequal samples do not establish distinct disease-area electrostatic windows.

\begin{table}[h]
\tbl{Charge-controlled binder design across three targets in BindCraft (filters
off; per-target $\beta$ grid, key settings shown). ``maximum'' is the maximum interface
pTM at that setting; ``$\ge0.5$'' is the fraction of designs reaching it.}
{\begin{tabular}{@{}llrrrr@{}}\toprule
Target & Setting & $n$ & net charge & maximum i\_pTM & \% $\ge0.5$\\ \colrule
PD-L1 & $\beta=0$ (vanilla)   & $28$ & $-5.8$  & $0.85$ & $39\%$\\
PD-L1 & $\beta=0.15$          & $24$ & $-2.1$  & $0.83$ & $62\%$\\
PD-L1 & $\beta=0.3$ (neutral) & $16$ & $-0.5$  & $0.83$ & $38\%$\\
PD-L1 & $\beta=2.0$ (poly-K)  & $16$ & $+66.9$ & $0.10$ & $0\%$\\
\colrule
IL-7R$\alpha$ & $\beta=0$ (vanilla) & $16$ & $-3.0$  & $0.33$ & $0\%$\\
IL-7R$\alpha$ & $\beta=0.5$         & $16$ & $+3.3$  & $0.79$ & $44\%$\\
IL-7R$\alpha$ & $\beta=1.0$         & $16$ & $+36.2$ & $0.29$ & $0\%$\\
\colrule
RBD & $\beta=0$ (vanilla) & $16$ & $-5.1$ & $0.32$ & $0\%$\\
RBD & $\beta=0.5$         & $16$ & $+7.2$ & $0.12$ & $0\%$\\ \botrule
\end{tabular}}
\label{sms:tbl-bindcraft}
\end{table}

\subsection{Charge control changes computed charge-window and surface-patch
proxies}
Under the chosen computational criteria, $0$ of $28$ PD-L1 baseline designs were
both inside the pI/charge window and above i\_pTM $0.5$, compared with $5$ of $24$
designs ($20.8\%$) at $\beta=0.15$. The overall i\_pTM-threshold hit rates were
$39.3\%$ and $62.5\%$, respectively, with overlapping bootstrap intervals. This
is enrichment under pre-specified in silico proxies; it does not demonstrate
experimental developability or binding, and it should be interpreted only as a
count within this finite sample.

\begin{figure}[h]
\centerline{\includegraphics[width=3.4in]{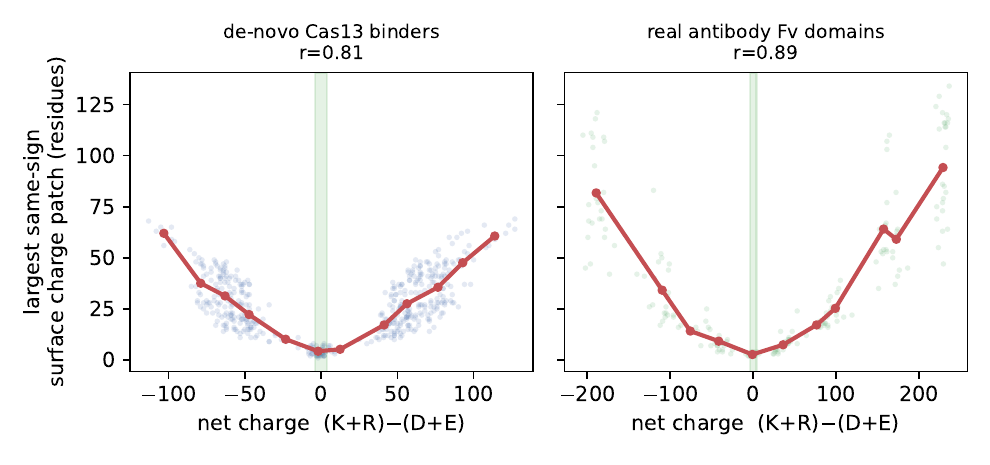}}
\caption{Reported association between absolute net-charge magnitude and the
largest connected same-sign residue set under a fixed-backbone C$\alpha$-neighbour
proxy for Cas13 designs ($r=0.81$) and antibody Fv designs ($r=0.89$). Points are
designs and the line is the binned mean. The proxy is not an
electrostatic-potential surface or an experimental developability measurement;
uncertainty should account for repeated designs from the same protein.}
\label{sms:fig-devspatial}
\end{figure}

Net charge does not encode spatial charge distribution. Using the fixed-backbone
C$\alpha$-neighbour proxy defined for Fig.~\ref{sms:fig-devspatial}, the reported
analysis associates larger absolute net-charge magnitude with a larger connected
same-sign residue set, with reported Pearson $r=0.81$ for Cas13 and $r=0.89$ for
antibody Fv domains. This establishes co-variation with the chosen geometric proxy
only. It does not show that neutralisation improves viscosity, polyspecificity or
aggregation, which were not measured, and it is not an electrostatic-potential
calculation.

In a separate analysis of $25$ antibody Fv domains, the manuscript reports that
$17$ of $25$ native sequences were within the selected $0$--$6$ charge range and
that the secant controller had MAE $3.8$ versus $6.4$ for a matched global bias.
These point estimates are consistent with lower error in this small Fv set, but no
clustered uncertainty is reported here. The result does not independently validate
the charge window or demonstrate the spatial-patch and experimental developability
claims.

\subsection{Robustness and a surface-restricted variant}
Exploratory sensitivity checks reported Cas13 secant-loop MAE of $6.4$, $5.6$ and
$5.6$ at sampling temperatures $0.1$, $0.2$ and $0.3$. These similar point
estimates do not establish temperature invariance without uncertainty and complete
output files. A surface-restricted bias was also reported to produce MAE $12.3$
versus $12.3$ for the uniform bias on RCSB and $4.4$ versus $7.9$ on Cas13, while
retaining about $55\%$ of the reachable charge range. This supports a possible
structural-footprint trade-off, but not a general claim that surface restriction is
at least as precise, because no paired uncertainty is reported and the underlying
result CSVs were not present in the audited snapshot.

\section{Limitations}\label{sms:limitations}
We report these results as a computational capability demonstration.
(i)~Isolated-chain self-consistency does not measure a complex interface. In the
$52$-complex seed-$0$ AlphaFold2-Multimer analysis, full-cohort fold and ipSAE
metrics declined at all nonzero settings, while DockQ declined clearly only at
$\pm3$. The independent-seed replication selected eight complexes using their
seed-$0$, $\beta=0$ DockQ and therefore cannot estimate a population effect for all
$52$ complexes. Its fold-tolerant subsets contained zero to four complexes and
gave metric-dependent results, preventing a general interface-preservation claim.
(ii)~The BindCraft sweeps use small, unequal samples and predicted metrics only;
PD-L1 success intervals overlap, IL-7R$\alpha$ is non-monotonic and RBD has no
strong design at any setting.
(iii)~$\beta$ is a temperature- and size-relative logit shift, so achieved charge
change is more interpretable than $\beta$ alone.
(iv)~The charge windows are antibody-derived heuristics and are not validated
cut-offs for small de novo binders.
(v)~Fig.~\ref{sms:fig-devspatial} uses a fixed-backbone C$\alpha$-neighbour proxy
rather than electrostatic potential or a measured developability endpoint;
row-level outputs and protein-clustered uncertainty should be released.
(vi)~The temperature, surface-restriction and Fv analyses require releasable
outputs and uncertainty estimates.
(vii)~All evidence is computational, and wet-lab validation remains necessary.

\section*{Code and data availability}
The charge-control and per-protein calibration scripts, the spatial-developability and
antibody-Fv analyses, and the charge-controlled BindCraft notebook (including the
amino-acid restype-indexing guardrails described in Methods) will be released
publicly upon publication. The AlphaFold2-Multimer prediction tables, independent
seed replication outputs, native-reference DockQ scores, paired changes,
cluster-bootstrap summaries and provenance manifest accompany the analysis branch.

\section{Conclusion}\label{sms:conclusion}
Building on ProteinMPNN's bias inputs, ZetaDial adds a post-sampling, per-protein
secant controller for net charge. Among the per-protein feedback controllers tested
it reduced mean charge-hit error on both datasets, and relative to a matched global
bias it reduced error on the heterogeneous RCSB set while remaining statistically
indistinguishable on Cas13. The complex-aware analysis characterises a foldability
cost that grows with $|\beta|$: native-reference DockQ shows clear interface-structure
loss at $\pm3$, and a selected-seed replication indicates that moderate settings can
also carry substantial, target-dependent interface costs, though it does not estimate
the effect across all $52$ complexes. The BindCraft, Fv and surface-patch analyses
are exploratory computational studies. The data therefore support a bounded claim of
improved mean control under heterogeneous response, not an unqualified superiority
claim or an experimental developability improvement.

\bibliographystyle{ws-procs11x85}
\bibliography{references}

\end{document}